\RequirePackage{fix-cm}
\documentclass[smallextended]{svjour3}
\smartqed  
\usepackage[table,xcdraw]{xcolor}
\usepackage{mathptmx}
\usepackage[utf8]{inputenc}
\usepackage{soul}
\usepackage{graphicx}
\usepackage{nicefrac}
\usepackage{amsfonts}
\usepackage[T1]{fontenc}
\usepackage{xspace}
\usepackage{flushend}
\usepackage{ifthen}
\PassOptionsToPackage{hyphens}{url}
\usepackage{url}
\usepackage{multirow}
\usepackage{listliketab}
\usepackage{amssymb}
\usepackage[linesnumbered,lined,boxed,commentsnumbered]{algorithm2e}
\usepackage{algorithmic}
\usepackage{booktabs}
\usepackage{comment}
\usepackage{amsmath}
\usepackage{bbding}
\usepackage{listings}
\usepackage{float}
\usepackage{placeins}
\usepackage[misc]{ifsym}
\usepackage{tabularx}
\usepackage{rotating}
\usepackage{natbib}
\usepackage{ulem}
\usepackage{censor}
\usepackage{pgfplots}
\pgfplotsset{compat=1.14}
\usepackage{pgfplotstable}
\usepackage{wrapfig}% biography <<<<<<<<<<  
\usepackage[hidelinks]{hyperref}
\usepackage{tablefootnote}
\definecolor{Fork}{HTML}{99CCFF}
\definecolor{Non-Fork}{HTML}{FFCCCB}

\definecolor{chestnut}{rgb}{0.8, 0.36, 0.36}
\definecolor{red}{rgb}{1, 0.25, 0.098}
\definecolor{blue}{rgb}{0.25, 0.44, 1}
\definecolor{purple}{rgb}{0.631, 0.475, 0.949}
\definecolor{pink}{rgb}{0.969, 0.341, 0.549}
\definecolor{navy}{rgb}{0.09, 0.10, 0.19}
\definecolor{light}{rgb}{0.71, 0.73, 0.77}
\definecolor{rqboxframe}{RGB}{70,130,180}
\makeatletter
\newsavebox{\rqbox@box}
\newlength{\rqbox@innerwd}
\newenvironment{rqbox}[1]{%
  \def\rqbox@title{#1}%
  \par\medskip\noindent
  \setlength{\fboxsep}{6pt}%
  \setlength{\fboxrule}{0.8pt}%
  \setlength{\rqbox@innerwd}{\dimexpr\linewidth-2\fboxsep-2\fboxrule\relax}%
  \begin{lrbox}{\rqbox@box}%
  \begin{minipage}{\rqbox@innerwd}%
  \setlength{\parindent}{0pt}%
  \setlength{\parskip}{0.35\baselineskip}%
  {\color{rqboxframe}\bfseries \rqbox@title}\par\vspace{2pt}%
}{%
  \end{minipage}%
  \end{lrbox}%
  \noindent\fcolorbox{rqboxframe}{gray!5}{\usebox{\rqbox@box}}%
  \par\medskip
}
\makeatother

\usepackage{tikz}

\colorlet{punct}{red!60!black}
\definecolor{background}{HTML}{EEEEEE}
\definecolor{delim}{RGB}{20,105,176}
\colorlet{numb}{magenta!60!black}

\lstdefinelanguage{json}{
    basicstyle=\small,
    numbers=left,
    numberstyle=\scriptsize,
    stepnumber=1,
    numbersep=8pt,
    showstringspaces=false,
    breaklines=true,
    frame=lines,
    backgroundcolor=\color{background},
    literate=
     *{0}{{{\color{numb}0}}}{1}
      {1}{{{\color{numb}1}}}{1}
      {2}{{{\color{numb}2}}}{1}
      {3}{{{\color{numb}3}}}{1}
      {4}{{{\color{numb}4}}}{1}
      {5}{{{\color{numb}5}}}{1}
      {6}{{{\color{numb}6}}}{1}
      {7}{{{\color{numb}7}}}{1}
      {8}{{{\color{numb}8}}}{1}
      {9}{{{\color{numb}9}}}{1}
      {:}{{{\color{punct}{:}}}}{1}
      {,}{{{\color{punct}{,}}}}{1}
      {\{}{{{\color{delim}{\{}}}}{1}
      {\}}{{{\color{delim}{\}}}}}{1}
      {[}{{{\color{delim}{[}}}}{1}
      {]}{{{\color{delim}{]}}}}{1},
}
\usepackage{makecell}
\usepackage{pifont}
\usepackage{epigraph} 
\usepackage[flushleft]{threeparttable}
\definecolor{backcolour}{rgb}{0.95,0.95,0.92}
\UseRawInputEncoding
\usepackage{fontawesome}

\definecolor{celestialblue}{rgb}{0.29, 0.59, 0.82}
\definecolor{awesome}{rgb}{0.0, 0.2, 0.6}
\definecolor{coolblack}{rgb}{0.0, 0.18, 0.39}
\definecolor{maroon}{cmyk}{0, 0.87, 0.68, 0.32}
\definecolor{halfgray}{gray}{0.55}
\definecolor{ipython_frame}{RGB}{207, 207, 207}
\definecolor{ipython_bg}{RGB}{247, 247, 247}
\definecolor{ipython_red}{RGB}{186, 33, 33}
\definecolor{ipython_green}{RGB}{0, 128, 0}
\definecolor{ipython_cyan}{RGB}{64, 128, 128}
\definecolor{ipython_purple}{RGB}{170, 34, 255}
\usepackage[strict]{changepage}
  \definecolor{ABlue}{HTML}{127bca}
 \definecolor{LHScolor}{HTML}{555555}
\usepackage{framed}

\definecolor{formalshade}{rgb}{1.0,1.0,1.0}
\definecolor{side}{rgb}{0.0,0.2,0.6}

\definecolor{gray(x11gray)}{rgb}{0.75, 0.75, 0.75}

\lstdefinelanguage{python}{
    morekeywords={access,and,break,class,continue,def,del,elif,else,except,exec,finally,for,from,global,if,import,in,is,lambda,not,or,pass,print,raise,return,try,while},
    morekeywords=[2]{abs,all,any,basestring,bin,bool,bytearray,callable,chr,classmethod,cmp,compile,complex,delattr,dict,dir,divmod,enumerate,eval,execfile,file,filter,float,format,frozenset,getattr,globals,hasattr,hash,help,hex,id,input,int,isinstance,issubclass,iter,len,list,locals,long,map,max,memoryview,min,next,object,oct,open,ord,pow,property,range,raw_input,reduce,reload,repr,reversed,round,set,setattr,slice,sorted,staticmethod,str,sum,super,tuple,type,unichr,unicode,vars,xrange,zip,apply,buffer,coerce,intern},
    sensitive=true,
    morecomment=[l]\#,
    morestring=[b]',
    morestring=[b]",
    morestring=[s]{'''}{'''},
    morestring=[s]{"""}{"""},
    morestring=[s]{r'}{'},
    morestring=[s]{r"}{"},
    morestring=[s]{r'''}{'''},
    morestring=[s]{r"""}{"""},
    morestring=[s]{u'}{'},
    morestring=[s]{u"}{"},
    morestring=[s]{u'''}{'''},
    morestring=[s]{u"""}{"""},
    literate=
    {á}{{\'a}}1 {é}{{\'e}}1 {í}{{\'i}}1 {ó}{{\'o}}1 {ú}{{\'u}}1
    {Á}{{\'A}}1 {É}{{\'E}}1 {Í}{{\'I}}1 {Ó}{{\'O}}1 {Ú}{{\'U}}1
    {à}{{\`a}}1 {è}{{\`e}}1 {ì}{{\`i}}1 {ò}{{\`o}}1 {ù}{{\`u}}1
    {À}{{\`A}}1 {È}{{\'E}}1 {Ì}{{\`I}}1 {Ò}{{\`O}}1 {Ù}{{\`U}}1
    {ä}{{\"a}}1 {ë}{{\"e}}1 {ï}{{\"i}}1 {ö}{{\"o}}1 {ü}{{\"u}}1
    {Ä}{{\"A}}1 {Ë}{{\"E}}1 {Ï}{{\"I}}1 {Ö}{{\"O}}1 {Ü}{{\"U}}1
    {â}{{\^a}}1 {ê}{{\^e}}1 {î}{{\^i}}1 {ô}{{\^o}}1 {û}{{\^u}}1
    {Â}{{\^A}}1 {Ê}{{\^E}}1 {Î}{{\^I}}1 {Ô}{{\^O}}1 {Û}{{\^U}}1
    {œ}{{\oe}}1 {Œ}{{\OE}}1 {æ}{{\ae}}1 {Æ}{{\AE}}1 {ß}{{\ss}}1
    {ç}{{\c c}}1 {Ç}{{\c C}}1 {ø}{{\o}}1 {å}{{\r a}}1 {Å}{{\r A}}1
    {€}{{\EUR}}1 {£}{{\pounds}}1
    {^}{{{\color{ipython_purple}\^{}}}}1
    {=}{{{\color{ipython_purple}=}}}1
    {+}{{{\color{ipython_purple}+}}}1
    {*}{{{\color{ipython_purple}$^\ast$}}}1
    {/}{{{\color{ipython_purple}/}}}1
    {+=}{{{+=}}}1
    {-=}{{{-=}}}1
    {*=}{{{$^\ast$=}}}1
    {/=}{{{/=}}}1,
    literate=
    *{-}{{{\color{ipython_purple}-}}}1
     {?}{{{\color{ipython_purple}?}}}1,
    identifierstyle=\color{black}\ttfamily,
    commentstyle=\color{ipython_cyan}\ttfamily,
    stringstyle=\color{ipython_red}\ttfamily,
    keepspaces=true,
    showspaces=false,
    showstringspaces=false,
    rulecolor=\color{ipython_frame},
    numberstyle=\tiny\color{halfgray},
    backgroundcolor=\color{ipython_bg},
    basicstyle=\scriptsize,
    keywordstyle=\color{ipython_green}\ttfamily,
}

\usepackage{datenumber}

\definecolor{chestnut}{rgb}{0.8, 0.36, 0.36}

\definecolor{chestnut}{rgb}{0.8, 0.36, 0.36}

\newcounter{dateone}\newcounter{datetwo}%
\newcommand{\daydifftoday}[3]{%
\setmydatenumber{dateone}{\the\year}{\the\month}{\the\day}%
\setmydatenumber{datetwo}{#1}{#2}{#3}%
\addtocounter{datetwo}{-\thedateone}%
\thedatetwo
}

\usepackage[skins,breakable]{tcolorbox}

\definecolor{Large}{HTML}{696969}
\definecolor{Negligible}{HTML}{D3D3D3}
\definecolor{Medium}{HTML}{808080}
\definecolor{Small}{HTML}{A9A9A9}

\journalname{Empirical Software Engineering}

\begin{document}

\title{An Exploratory Study of Dependabot Cooldown Adoption in Open-Source GitHub Projects}

\author{Hidetake Tanaka \and
        Rikuto Tsuchida \and
        Kazumasa Shimari \Letter \and
        Raula Gaikovina Kula \and
        Kenichi Matsumoto
}

\institute{
    \Letter~Corresponding author - Kazumasa Shimari \\
    Hidetake Tanaka \at Nara Institute of Science and Technology, Japan \\
    \email{tanaka.hidetake.te0@naist.ac.jp} \and
    Rikuto Tsuchida \at Nara Institute of Science and Technology, Japan \\
    \email{tsuchida.rikuto.tq5@naist.ac.jp} \and
    Kazumasa Shimari \at Wakayama University, Japan \\
    \email{shimari@wakayama-u.ac.jp} \and
    Raula Gaikovina Kula \at The University of Osaka, Japan \\
    \email{raula-k@ist.osaka-u.ac.jp} \and
    Kenichi Matsumoto \at Nara Institute of Science and Technology, Japan \\
    \email{matumoto@is.naist.jp}
}

\date{Received: date / Accepted: date}
\setstcolor{red}
\maketitle

% 150-250 words
\begin{abstract}
Automated dependency updates can rapidly propagate malicious package releases before maintainers and the broader community have enough time to detect them.
In July 2025, GitHub made \texttt{Dependabot cooldown} generally available as a defense against software supply chain attacks.
However, the effects of its early adoption remain unknown.
In this exploratory study, we empirically examine how popular open-source GitHub repositories adopt and configure the feature and investigate their motivations.
We find that security concerns motivated 83 of 92 adoption events with known motivations. Security linter warnings triggered 43 of 75 security-only adoptions.
Among 251 ecosystems within repositories that retained cooldown, 97.2\% set a general delay. Of these, 64.3\% used seven days, while use of each update type setting was below 10\%.
Early adopters therefore favor simple default delays over fine-grained controls. These findings suggest that tools could provide robust defaults reflecting ecosystem support and reserve fine-grained controls for dependencies with clear update priorities.

\keywords{software supply chain \and dependency management \and open-source security}
\end{abstract}

\section{Introduction}

Software supply chain attacks have emerged as an important threat to open-source ecosystems~\citep{ladisa2023taxonomy, ohm2020backstabber}.
Attackers can publish malicious packages to registries or compromise existing ones, affecting numerous downstream projects through dependency chains~\citep{zimmermann2019small, duan2021towards}.
These attacks exploit the trust that developers place in package ecosystems and their automated dependency management workflows~\citep{zahan2022weak, williams2025research}.
Recent incidents such as the XZ Utils backdoor and the protestware cases in npm show that even updates published by legitimate maintainers can carry malicious payloads~\citep{przymus2025wolves, kula2022war}.
The time window between the publication of a new package version and its adoption by developers represents an important attack vector~\citep{he2025pinning, ohm2020backstabber}.
Dependency management bots such as Dependabot and Renovate automate the update process~\citep{erlenhov2022dependency, alfadel2021dependabot}, but they can also automate the rapid adoption of malicious packages if no safeguard is in place.
\citet{he2023automating} report that 56.2\% of update-check frequencies declared in Dependabot configurations are set to daily, and that 8.17\% of Dependabot pull requests are merged by third-party auto-merge implementations such as continuous integration workflows or GitHub Apps. Such practices may shorten the time between release and adoption, reducing the time available for the community to detect malicious packages before they are adopted.

To address this risk, several dependency management bots and package managers have introduced cooldown features, which are configurable delay periods before a newly published package version is automatically adopted.
Among these, GitHub's Dependabot released its cooldown configuration in July 2025, positioning it as a supply chain security feature and describing four intended benefits: reduced update noise, responsiveness to critical security patches, granular control through separate delays for each Semantic Versioning (SemVer) level, and flexible scheduling that integrates with existing update intervals.\footnote{\url{https://github.blog/changelog/2025-07-01-dependabot-supports-configuration-of-a-minimum-package-age/}}
Unlike cooldown options built into package managers (available in npm, pnpm, Yarn, Bun, and uv), Dependabot's implementation supports a wide range of package ecosystems (npm, pip, Maven, Gradle, Bundler, Cargo, Composer, NuGet, Go modules, Docker, GitHub Actions, Terraform, and more) under a single configuration schema.

Despite the availability of Dependabot's cooldown feature and these stated benefits, empirical evidence remains limited on how widely it is adopted or how it is configured in practice.
Understanding who adopts the feature, and how they configure and evolve it, provides the empirical basis needed for future effectiveness studies and for tool-design decisions.

In this study, we conduct an exploratory empirical analysis of early Dependabot cooldown adoption among top-starred open-source GitHub repositories and provide a focused characterization of early adopters' configurations.
By focusing on a single tool that supports multiple package ecosystems under a unified schema, we avoid cross-tool heterogeneity while enabling cross-ecosystem comparison.
Specifically, we address the following three research questions:

\begin{itemize}
\item \textbf{RQ1 (Adoption):} How do projects adopt and abandon Dependabot cooldown?
\item \textbf{RQ2 (Adopters):} What distinguishes Dependabot cooldown adopters?
\item \textbf{RQ3 (Configuration):} How do adopters configure and adjust Dependabot cooldown?
\end{itemize}

The main contributions of this study are as follows:
\begin{itemize}
\item Empirical evidence on Dependabot's cooldown feature, characterizing early adoption among popular open-source GitHub repositories across the diverse package ecosystems it supports under a single configuration schema.
\item A combined quantitative and qualitative investigation of who adopts the feature, including project characteristics and adoption motivations through analysis of related Issues and Pull Requests.
\item Evidence from the 135 observed early adopters on how they use the configuration mechanisms associated with Dependabot's stated benefits (e.g., SemVer-level granular control, flexible scheduling), with implications for tool designers and open-source maintainers.
\end{itemize}

\section{Related Work}
\label{sec:related}

\subsection{Software Supply Chain Security}

Software supply chain attacks have attracted increasing attention from both researchers and practitioners.
\citet{ladisa2023taxonomy} presented a comprehensive taxonomy of attacks on open-source software supply chains, categorizing attack vectors across the entire software development lifecycle.
Their taxonomy identifies the publication of malicious packages as a prevalent attack strategy, where attackers exploit the trust inherent in package registries.

\citet{ohm2020backstabber} collected and analyzed real-world instances of open-source supply chain attacks, providing a systematic overview of attack types and techniques.
Their work highlighted that attackers frequently target package registries by publishing packages with names similar to popular ones.

\citet{zimmermann2019small} studied security threats in the npm ecosystem and demonstrated that the heavy reliance on a small number of highly influential packages creates widespread risks.
Their findings showed that compromising a single popular package could affect thousands of downstream projects.

\citet{duan2021towards} measured supply chain attacks on package managers for interpreted languages, quantitatively assessing the scale and impact of such attacks.
\citet{vu2020typosquatting} investigated typosquatting and combosquatting attacks on the Python ecosystem, revealing that these attacks exploit common typographical errors in package names to distribute malicious code.

Recent high-profile incidents further show that legitimate maintainers and the update path itself can become the attack surface.
The 2024 XZ Utils backdoor was planted after a long social-engineering campaign that gained the trust of the project's maintainer~\citep{przymus2025wolves, lins2025killchain}.
Protestware cases such as \texttt{colors}, \texttt{faker}, and \texttt{node-ipc} demonstrated that even legitimate maintainers can ship destructive updates in protest of world events~\citep{kula2022war, cheong2024ethical}.
Industry reports likewise document a sustained rise in malicious packages published to open-source registries~\citep{sonatype2026supplychain}, and \citet{williams2025research} synthesize these developments into a research roadmap for software supply chain security.
These cases share a common pattern: the malicious payload enters as an ordinary version release and reaches the downstream projects that adopt it quickly.

Beyond characterizing attacks, researchers have also proposed techniques to detect malicious packages automatically.
\citet{sejfia2022practical} developed an automated approach that identifies malicious npm packages based on package metadata and code features, and \citet{guo2023empirical} conducted an empirical study of malicious code in the PyPI ecosystem.
\citet{gonzalez2021anomalicious} detect anomalous and potentially malicious commits from commit logs and repository metadata, and \citet{vu2021lastpymile} identify discrepancies between the source repository of a package and the artifact distributed through the registry.
Complementing these detection approaches, \citet{zahan2022weak} identified weak-link signals in the npm supply chain, such as expired maintainer domains and install scripts, that indicate a heightened risk of compromise.
These detection techniques are complementary to cooldown mechanisms: cooldown provides the time window during which detection tools and community reports can identify malicious packages before downstream adoption.

While these studies have advanced our understanding of attack characteristics, their impact, and detection techniques, they leave the adoption of defense mechanisms such as cooldown features insufficiently understood.
Our study addresses this gap by empirically investigating how cooldown features are configured and adopted across ecosystems.

\subsection{Dependency Management Bots}

Dependency management bots automate the process of keeping library dependencies up to date and have become increasingly popular in open-source software development~\citep{he2023automating}.

\citet{alfadel2021dependabot} investigated the use of Dependabot security pull requests, analyzing how developers respond to automated security updates.
Their study revealed that while Dependabot helps developers address known vulnerabilities, a significant portion of security pull requests remain unmerged.

\citet{mirhosseini2017automated} found that projects using automated pull request notifications updated dependencies 1.6 times as often as projects that did not use notification tools, although only about one third of the automated pull requests were merged.
Their survey identified breaking changes, difficulty understanding the implications of updates, and migration effort as developers' principal concerns.

The benefits of dependency bots come with a notification cost.
\citet{wessel2018power} characterized the widespread use of bots in open-source projects, and \citet{wessel2021disturb} identified the interruptions and noise that bot interactions impose on maintainers as a central challenge.
\citet{rombaut2023greenkeeper} quantified this overhead for the Greenkeeper dependency bot in npm, showing that a large share of its notifications demand maintainer attention without delivering a corresponding benefit.

These studies have examined the prevalence, effectiveness, and overhead of dependency management bots, but evidence remains limited on the usage of cooldown configurations offered by these bots.
Because cooldown delays and batches updates, it also bears on this overhead by reducing the pull requests that maintainers must triage, yet how projects configure such delay-based safeguards remains unclear.
Our work complements this line of research by specifically examining how projects configure them across ecosystems.

\subsection{Dependency Update Lags}

Dependency update lag refers to the delay between the release of a new library version and its adoption by downstream projects.
\citet{kula2018developers} conducted an empirical study on whether developers update their library dependencies in response to security advisories.
Their findings indicated that many developers are unaware of or reluctant to update their dependencies, even when security advisories are available.

\citet{chinthanet2021lags} investigated the lags in the release, adoption, and propagation of npm vulnerability fixes.
Their study revealed that significant delays exist at each stage of the vulnerability fix pipeline, from the release of a patch to its propagation through the dependency network.

The notion of technical lag, how far a deployment is behind the latest available releases, was introduced by \citet{gonzalezbarahona2017technical} and formalized and applied to npm by \citet{zerouali2019formal}.
\citet{cox2015measuring} showed that low dependency freshness is associated with security risk, which highlights the cost side of delaying updates.
Release and update dynamics also differ substantially across ecosystems: \citet{decan2019empirical} compared the evolution of dependency networks in seven package ecosystems, and \citet{wittern2016look} documented the high release frequency and dense dependency structure of the npm ecosystem.
\citet{cogo2021downgrades} further showed that developers respond to problematic releases by downgrading dependencies, which suggests that adopting a release immediately carries a rollback cost that a waiting period can avoid.

Dependabot's separate cooldown delays for each SemVer level presuppose the compatibility conventions of semantic versioning.
Empirical studies temper this assumption: \citet{raemaekers2017semantic} and \citet{ochoa2022breaking} showed that breaking changes also appear in minor and patch releases in the Maven ecosystem, and \citet{bogart2016break} found that ecosystems differ in their practices and values around breaking changes.
\citet{decan2021semver} showed that most dependency constraints follow semver-compatible ranges that accept new releases automatically, which is precisely the mechanism that lets a malicious release propagate quickly.
The way developers declare version constraints also shapes their exposure: \citet{dietrich2019versioning} documented the diversity of versioning practices across ecosystems, and \citet{jafari2022smells} identified constraint choices such as restrictive pinning and wildcards as dependency smells with maintenance and security consequences.

It is important to distinguish between cooldown features and unintentional update lags.
Cooldown introduces a controlled, intentional delay designed to protect against supply chain attacks, whereas update lags arise from developer inattention or reluctance to update dependencies.
Our study focuses on the former, that is, on the deliberate adoption and configuration of cooldown mechanisms as a proactive defense strategy.

Cooldown as a concept predates Dependabot's 2025 release.
Renovate introduced its \texttt{stabilityDays} option in 2019 to address a stability concern: npm allows packages to be unpublished within 72 hours of release\footnote{\url{https://docs.npmjs.com/policies/unpublish}}, and immediately adopting new versions could break builds when packages were retracted.
In contrast, the wave of cooldown features introduced in 2025 and 2026 by Dependabot and by package managers such as pnpm, Yarn, Bun, npm, and uv was explicitly motivated by supply chain security, aiming to prevent the rapid adoption of malicious packages amid a rising threat of malicious package updates~\citep{sonatype2026supplychain, williams2025research}.
Our study focuses on Dependabot as one instance of this newer, security-driven wave.

\section{Study Design}
\label{sec:study-design}

\subsection{Overview}

We conduct an exploratory empirical study of the early adoption of Dependabot's cooldown feature across the package ecosystems that Dependabot supports, using top-starred open-source repositories hosted on GitHub.
The study consists of three phases. First, identification of the target tool, second, data collection from GitHub repositories, and third, quantitative and qualitative analysis to answer our three research questions.

We limit the study to Dependabot, GitHub's official dependency management bot, which made its cooldown configuration generally available in July 2025 and accounts for more than 65\% of dependency management activity on GitHub~\citep{rebatchi2024dependabot}. We select it because it provides a unified configuration schema across its supported package ecosystems and offers fine-grained settings, including separate delays for each SemVer level and scoping at the package level, that allow us to compare configuration choices across ecosystems and repositories.

\subsection{Data Collection}
\label{sec:data-collection}

\begin{figure*}[tbp]
\centering
\includegraphics[width=\textwidth]{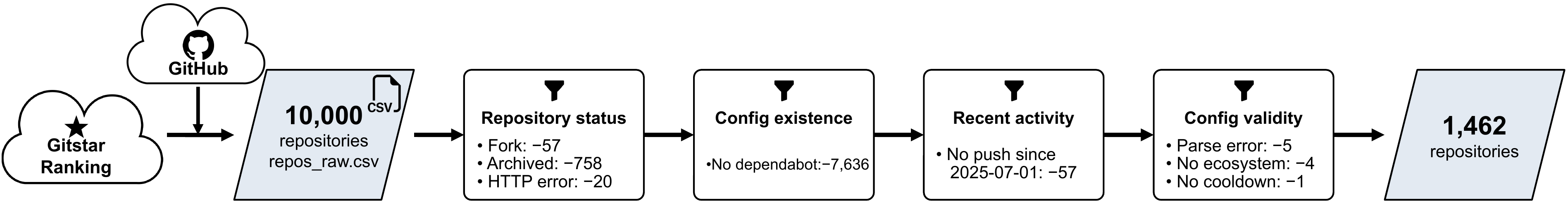}
\caption{Filtering stages from the candidate repositories to the analysis population, with the number of repositories excluded at each stage}
\label{fig:dataset-pipeline}
\end{figure*}

Figure~\ref{fig:dataset-pipeline} summarizes how we narrow the candidate repositories down to the analysis population. We first retrieve candidate repositories. We then filter them by repository status, Dependabot configuration existence, recent activity, and configuration validity.

During candidate retrieval, we build the initial set from the 10,000 GitHub repositories with the most stars listed in Gitstar Ranking,\footnote{\url{https://gitstar-ranking.com/repositories}} a publicly available ranking service that aggregates GitHub's public data. By restricting the candidates to repositories with the most stars, we focus the analysis population on popular and visible projects where user bases, dependency management activity, and security practices are more likely to be observable, making early adoption of a new feature more likely to appear in the data.

We retrieved the list on 2026-05-01 12:09 UTC, roughly 12 hours after the study cutoff of 2026-04-30 24:00 UTC. Gitstar Ranking refreshes its data on a fixed schedule, so the ranking we observed may reflect star counts that lag by up to 7 days.

When checking repository status, we use GitHub REST API metadata to exclude 57 forked and 758 archived repositories. This avoids known pitfalls in GitHub mining studies~\citep{kalliamvakou2014promises}. We also exclude 20 repositories whose metadata we could not retrieve. This stage leaves 9{,}165 repositories.

When checking for configuration files, we query the GitHub REST API for both paths that Dependabot officially recognizes, \path|.github/dependabot.yml| and \path|.github/dependabot.yaml|, and we refer to both variants as the Dependabot configuration file. In these YAML files, each entry directly under the \texttt{updates} field identifies a target ecosystem and directory, and cooldown is configured within such an entry. We exclude 7{,}636 repositories that contain neither file and retain the 1{,}529 repositories that contain at least one variant.

When filtering for recent activity, we retain a repository if its \texttt{pushed\_at}, the time of its last change, is on or after 2025-07-01, the date when Dependabot announced the general availability (GA) of the cooldown feature. Every repository in our observation therefore had the opportunity to adopt cooldown. This filter removes 57 repositories and leaves 1{,}472. We set the observation cutoff at 2026-04-30 to capture one month of activity after the 2026-03-31 axios incident, in which malicious axios versions were published from a compromised npm account and injected a dependency that installed a remote access trojan~\citep{axiosPostmortem2026}. The observation window therefore spans roughly ten months from 2025-07-01 to 2026-04-30.

When validating configurations, we enumerate every commit within the observation window that touches the Dependabot configuration file, take a full YAML snapshot at each such commit, and parse the snapshot. We exclude five repositories whose configuration at the cutoff fails to parse due to YAML syntax errors such as incorrect indentation or missing required headers. We further exclude four repositories whose \texttt{package-ecosystem} field is empty across every \texttt{updates} entry at the cutoff. An empty \texttt{package-ecosystem} field indicates the default GitHub-generated template used without customization. Because such a file configures no ecosystem, we do not treat it as a valid Dependabot configuration.

During our observation window, Dependabot took the delay from \texttt{default-days} and the three \texttt{semver-*-days} keys and treated an absent key as zero days\footnote{\url{https://github.com/dependabot/dependabot-core/blob/2eaf27b6c3882b791f6feab1ddcd933c8d63c4a7/common/lib/dependabot/package/release_cooldown_options.rb\#L22-L25}}. One repository uses the unrecognized \texttt{default: 4} instead of \texttt{default-days: 4} in its cooldown. Because none of the recognized day keys is present, its cooldown delays no update even though it appears in the file. We therefore exclude the repository from the analysis population before classifying adoption status. After these steps, the final analysis population consists of 1{,}462 repositories.

To characterize both the uptake and persistence of cooldown, we classify each repository into one of four mutually exclusive adoption states based on its configuration history during the observation window and its status at the cutoff.
\begin{enumerate}
\item \textbf{Non-adopter.} The repository never held a cooldown during the observation window.
\item \textbf{Continuous adopter.} The repository adopted cooldown and still holds it at the cutoff without its cooldown count having dropped to zero after adoption.
\item \textbf{Re-adopted adopter.} The repository's cooldown count dropped to zero, later rose above zero again, and remains above zero at the cutoff.
\item \textbf{Abandoned adopter.} The repository adopted cooldown during the observation window but no longer holds it at the cutoff.
\end{enumerate}

We apply the same four states to individual ecosystems within repositories. If an ecosystem's \texttt{updates} entry is removed entirely, we treat it as a discontinuation of management for that ecosystem rather than abandonment of cooldown and exclude it from this classification.

For the repository-level analysis in RQ2, we combine adoption status derived from configuration histories with repository metadata, dependency counts, Dependabot pull request activity, and the history of configuration edits before GA. Table~\ref{tab:rq2-measures} defines the measures and their observation periods. The replication package documents the schema of each file, the joins, the API calls behind each measure, and how we handle the repositories for which a retrieval fails. For the repositories whose dependency graph is unavailable, we count dependencies with Bibliothecary\footnote{\url{https://github.com/librariesio/bibliothecary}}, a Ruby library that parses manifest and lock files.

\begin{table*}[tbp]
\caption{Repository-level measures collected for the analyses, their operational definitions, data sources, and measurement periods. GA denotes general availability.}
\label{tab:rq2-measures}
\centering
\scriptsize
\setlength{\tabcolsep}{3pt}
\renewcommand{\arraystretch}{1.08}
\begin{tabularx}{\textwidth}{@{}>{\raggedright\arraybackslash}p{0.16\textwidth}>{\raggedright\arraybackslash}X>{\raggedright\arraybackslash}p{0.30\textwidth}@{}}
\toprule
\textbf{Measure} & \textbf{Operational definition} & \textbf{Source and period} \\
\midrule
Adoption status & Whether a repository held at least one cooldown at any point in the observation window. & Dependabot configuration file history over the observation window. \\
Stars & Number of users who starred the repository. & GitHub repository metadata at the cutoff. \\
Repository age & Years from the repository creation date to the observation cutoff. & GitHub repository metadata at the cutoff. \\
Contributors & Number of contributors returned for the repository. & GitHub contributors endpoint with data collected for the cutoff. \\
Owner type & Whether the repository owner is a User or an Organization. & GitHub repository metadata at the cutoff. \\
Dependencies & Number of distinct pairs of ecosystem and package name, with different versions of one package counted once. & GitHub dependency graph at the cutoff, supplemented with Bibliothecary when unavailable. \\
Bot pull request count & Number of pull requests opened by \path|dependabot[bot]| for the repository. & Dependabot pull requests over the observation window. \\
Merge time & Per-repository median hours from creation to merge among merged Dependabot pull requests. & Dependabot pull requests over the observation window for repositories with at least one merged bot pull request. \\
Auto-merge & Whether GitHub's native auto-merge was enabled on at least one Dependabot pull request. & Dependabot pull requests over the observation window. \\
\texttt{SECURITY.md} & Whether the repository published a \path|SECURITY.md| file. & Repository files at the cutoff. \\
Primary language & Dominant programming language reported for the repository. & GitHub repository metadata in \texttt{repos} at the cutoff. \\
Pre-GA edit rate & Commits editing the Dependabot configuration from its first addition through GA, divided by the elapsed years. & Dependabot configuration file history before GA for repositories that held the file at GA. \\
\bottomrule
\end{tabularx}
\end{table*}

We use a pre-GA annual edit rate to examine whether active maintenance of a repository's Dependabot configuration is related to cooldown adoption. We restrict the measure to edits before GA because an edit count during the observation window would include the cooldown-introduction commit for adopters, thereby confounding maintenance activity with adoption itself. Among repositories that already held the configuration file at GA, we count the commits that edited the file from its first addition through GA and divide this count by the years in that span to obtain the annual edit rate. Of the 1{,}462 repositories, 1{,}289 had the file at GA with an identifiable first commit and enter this analysis. We exclude 154 that had no Dependabot configuration before GA, 18 with no commit before GA because they were created later or their history was rewritten, and 1 for which we could not retrieve the GA snapshot.

\subsection{Analysis Methods}

We describe the analysis methods for each research question below.
Following the conventional 0.05 level for statistical testing~\citep{fisher1925statistical}, we use $p < 0.05$ as the threshold for statistical significance in standalone tests.

\subsubsection{RQ1: Adoption and Abandonment}
To answer RQ1, we use the Dependabot configuration histories and related development artifacts constructed in Section~\ref{sec:data-collection}. We organize the analysis into four complementary steps.
\begin{enumerate}
\item \textbf{Cutoff snapshot.} We examine the distribution of adoption and abandonment at the cutoff.
\item \textbf{Repository-level timeline.} We examine the cumulative adoption timeline at the repository level.
\item \textbf{Ecosystem-level analysis.} We examine adoption rates across ecosystems.
\item \textbf{Motivation analysis.} We conduct a qualitative content analysis of the motivations behind adoption and abandonment.
\end{enumerate}

The first three steps are quantitative analyses based on the commit histories of Dependabot configuration files. The fourth uses manual coding of related commits, pull requests, and issues.

\paragraph{\textbf{Step 1: Cutoff Snapshot.}}
To gauge how widely cooldown has spread, we examine the overall breakdown of adoption and abandonment. For the cutoff distribution, we count both repositories and ecosystems within repositories under the four adoption states defined in Section~\ref{sec:data-collection}.

\paragraph{\textbf{Step 2: Repository-Level Timeline.}}
We also examine the repository-level adoption timeline. A repository's adoption timestamp is the commit at which a cooldown first appears anywhere in the repository. We plot the cumulative count of adopter repositories.

\paragraph{\textbf{Step 3: Ecosystem-Level Analysis.}}
Because adoption may differ across ecosystems, we also examine ecosystem-level adoption rates at the cutoff.
The unit for the adoption rate is also an ecosystem within a repository. For each ecosystem, we divide the number of repositories that hold a cooldown in it by the number of repositories that still declare it at the cutoff.

One repository can declare several ecosystems, and cooldown adoption may follow a policy that covers the whole repository rather than the traits of an individual ecosystem, so observations that share a repository are correlated. We therefore test differences in the adoption rate across ecosystems with a mixed-effects logistic regression~\citep{breslow1993approximate} that places a random intercept at the repository level. We compare a model with ecosystem as a fixed effect against a model with only the random intercept in a likelihood-ratio test and base our inference on the overall contribution of the ecosystem fixed effect. This test evaluates whether cooldown adoption propensity differs across ecosystems overall after accounting for repository-level variation.

Including ecosystems declared by very few repositories can make the estimates unstable. Our primary analysis covers all 28 ecosystems, and we also run a sensitivity check restricted to the ecosystems declared by at least 30 repositories. Quasi-complete separation for ecosystems with no observed adopters and limited within-repository contrasts from repositories that declare only one ecosystem can destabilize the individual fixed-effect estimates. The mixed-effects model restricted to ecosystems declared by at least 30 repositories did not converge. For this sensitivity analysis, we therefore fit a population-averaged logistic regression using generalized estimating equations (GEE)~\citep{liang1986longitudinal}, clustered by repository. The model assumes an exchangeable working correlation within repositories and uses robust sandwich standard errors.

\paragraph{\textbf{Step 4: Motivation Analysis.}}
For the fourth analysis, we manually classify the motivation behind each repository's first-time cooldown adoption event ($n=135$), following established guidance on qualitative content analysis and thematic synthesis~\citep{krippendorff2019content, cruzes2011thematic}. For each event, we inspect the commit message, the pull request title and body, and the title and body of any issue referenced by the commit or pull request.

Two authors independently assign one of the following labels to each of these 135 repository-level first-time adoption events.
\begin{enumerate}
\item Security (Linter). The cooldown was introduced in response to a warning from a linter that detects vulnerabilities, such as zizmor\footnote{\url{https://github.com/zizmorcore/zizmor}}, a static analysis tool for GitHub Actions workflows that flags security-related configuration issues.
\item Security (No-Linter). The introduction was motivated by supply chain security and was unrelated to any linter.
\item Maintenance. The goal was to reduce the volume of Dependabot update pull requests and to improve code maintainability, as in wording that calls Dependabot pull requests noisy or expresses a wish to receive fewer of them.
\item Security and maintenance. The motivation spans both aspects. This includes wording that explicitly mentions both maintenance and security, as well as broader wording about stability or about adopting upstream bug and regression fixes that fits either side.
\item Template tracking. The cooldown adoption pull request was created automatically by following a repository that manages shared configuration, without an explicit adoption decision in the adopting repository.
\item Unknown. No clear motivation is stated.
\end{enumerate}

We treat wording about stability or about upstream bugs and regressions as Security and maintenance, because such wording carries both a maintenance aspect and a security aspect. When a pull request body or commit message links to an external article, we also read the article and include it in this judgment.

We measure inter-rater reliability using Cohen's kappa~\citep{cohen1960coefficient} and interpret the agreement level following the guidelines of \citet{landis1977measurement}. We resolve disagreements through discussion and assign a single label. We report the distribution of the motivation categories together with representative examples for each.

We also study the motivations behind abandoning cooldown. At the repository level, we examine commits at which a repository's cooldown count drops to zero. Abandonment at the level of an ecosystem within a repository is also possible. For such an ecosystem, we examine the commits that remove cooldown from one ecosystem while the \texttt{updates} entry stays in place and the repository keeps cooldown on its other ecosystems. We exclude removals that delete the \texttt{updates} entry itself, which discontinue management of that ecosystem rather than reject cooldown, and we likewise exclude removals that delete the Dependabot configuration file entirely. We read the pull request body and the commit message that removed cooldown, and we analyze the stated motivations.

\subsubsection{RQ2: Adopter Characteristics}
Adoption prevalence alone does not show whether cooldown has diffused broadly or remains concentrated among projects with greater maintenance capacity, heavier dependency-update workloads, or more visible security practices. RQ2 therefore compares adopter and non-adopter repositories to identify characteristics associated with early adoption. This characterization clarifies which project populations are represented among early adopters and provides a basis for future tool design and adoption studies.

The unit of analysis for RQ2 is a repository. We define the 135 repositories that held at least one cooldown during the observation window as adopters and the 1{,}327 that never held one as non-adopters. Adoption status is the binary outcome, and the repository measures defined in Table~\ref{tab:rq2-measures} serve as the explanatory variables.

We select attributes that capture six dimensions that may relate to early adoption: project visibility and maturity, maintenance capacity and governance, the burden of dependency updates, update workflow, visible security practices, and technical context. Stars and repository age represent visibility and maturity. Contributors and owner type represent maintenance capacity and governance. Dependencies and bot pull request count represent update burden. Merge time, auto-merge, and the pre-GA edit rate represent update workflow. The presence of \path|SECURITY.md| represents visible security practice. Primary language represents technical context. Prior work reports that daily schedules and auto-merge are widespread in Dependabot use~\citep{he2023automating}, giving further reason to examine whether these operational characteristics differ between adopters and non-adopters.

The analysis proceeds in two stages. In the first stage, we conduct univariable comparisons between adopters and non-adopters using six continuous variables: stars, repository age, contributors, dependencies, bot pull request count, and merge time. We also compare three binary variables: owner type, \path|SECURITY.md| presence, and auto-merge use. Primary language is compared separately, and the pre-GA edit rate serves as a supplementary analysis. In the second stage, to adjust for correlation among attributes, we fit a multivariable logistic regression over the 1{,}461 repositories with a dependency count. The main model includes the five continuous variables other than merge time, the three binary variables, and primary language. Because merge time and the pre-GA edit rate are defined only for subsets, we add each separately to the main model in a sensitivity analysis.

In the main model, primary language is categorical, with Python, the most common individual language among adopters, as the reference. To stabilize its coefficients, we keep only languages with at least 10 adopter repositories as individual categories and fold the rest into Other. We apply a log1p transform to the continuous variables and standardize them to zero mean and unit standard deviation, so an odds ratio expresses the change in adoption odds per one standard deviation increase on the transformed scale.

For continuous variables, we use the Mann-Whitney U test~\citep{mann1947test} and report Cliff's delta~\citep{cliff1993dominance}. For binary variables, we use the chi-squared test~\citep{pearson1900criterion} and report Cram\'er's V~\citep{cramer1946mathematical}. Primary language has many cells with small expected counts, so we use Fisher's exact test on the full ungrouped distribution and report Cram\'er's V. Because exact enumeration is impractical, we approximate the p-value with 10{,}000 Monte Carlo resamples using a fixed seed of 0. In the paper and the multivariable model, we show languages with at least 10 adopter repositories individually and fold the rest into Other. The full distribution is included in the replication package.

To interpret the magnitude of Cliff's delta, we adopt the thresholds documented for the \texttt{cliff.delta} function in the R package \texttt{effsize}~\citep{torchianoEffsizeCliffDelta}, classifying the absolute effect size as negligible for |$\delta$| $<$ 0.147, small for 0.147 $\leq$ |$\delta$| $<$ 0.33, medium for 0.33 $\leq$ |$\delta$| $<$ 0.474, and large for |$\delta$| $\geq$ 0.474. Following \citet{cohen1988statistical}, we use small, medium, and large thresholds of 0.10, 0.30, and 0.50 for Cram\'er's V. Every contingency table in this comparison has two columns, adopters and non-adopters, so these thresholds for one degree of freedom apply throughout.

For the univariable comparisons used to identify adopter characteristics, we treat the six continuous-variable and three binary-variable comparisons as one family of nine tests. We control its false discovery rate at $q = 0.05$ with the Benjamini--Hochberg (BH) procedure~\citep{benjamini1995controlling}, base statistical significance on the BH-adjusted p-values, and report only these adjusted values.

For the multivariable model, we use Spearman rank correlations to describe pairwise associations among the five continuous and three binary repository attributes and calculate variance inflation factors (VIFs) for all explanatory variables in the main model, including primary language~\citep{dormann2013collinearity}. We use a VIF of 5 as an empirical screening criterion for further investigation~\citep{james2021introduction}, while treating it as a guide rather than a definitive cutoff~\citep{obrien2007caution}. We estimate the model by maximum likelihood and report an odds ratio (OR) adjusted for the other variables in the model, a 95\% confidence interval (CI), and a two-sided p-value for each explanatory variable. We treat coefficient-level inference as exploratory and interpret the estimates primarily using the odds ratios and confidence intervals. We report McFadden's pseudo $R^2$~\citep{mcfadden1974conditional} as a summary of model fit and the apparent area under the receiver operating characteristic curve (AUC)~\citep{hanley1982meaning} as a measure of in-sample discrimination.

Because the pull request metrics span the full observation window, adopter measurements may include post-adoption observations and should not be interpreted purely as pre-adoption characteristics. We discuss this limitation in Section~\ref{sec:discussion}.
\subsubsection{RQ3: Configuration Choices}
It is not yet known how adopters configure cooldown, how far they use fine-grained controls such as SemVer granularity and the include and exclude lists, and how they change it after adoption. To clarify this, in RQ3 we analyze the cooldown configuration values.
We organize the analysis into three parts: the set rate and value distribution of each key, value changes after adoption, and the comparison of these values within the same repository.

The keys available within a cooldown field are listed in Table~\ref{tab:cooldown-config}. The \texttt{default-days} key defines the baseline delay for the target ecosystem. For ecosystems that follow SemVer~\citep{decan2021semver}, \texttt{semver-major-days}, \texttt{semver-minor-days}, and \texttt{semver-patch-days} override \texttt{default-days} for their respective update types. The \texttt{include} and \texttt{exclude} lists use wildcard patterns to scope the dependencies to which cooldown applies, and \texttt{exclude} takes precedence when a dependency matches both lists.

\begin{table}[t]
\caption{Keys of the Dependabot cooldown configuration~\citep{githubDependabotOptions}. SemVer denotes Semantic Versioning.}
\label{tab:cooldown-config}
\centering
\setlength{\tabcolsep}{4pt}
\renewcommand{\arraystretch}{1.05}
\begin{tabular}{@{}lp{0.60\columnwidth}@{}}
\toprule
\textbf{Key} & \textbf{Description} \\
\midrule
\texttt{default-days} & Baseline delay for dependencies without a more specific rule. \\
\texttt{semver-major-days} & Delay for major SemVer updates. Takes precedence over \texttt{default-days}. \\
\texttt{semver-minor-days} & Delay for minor SemVer updates. Takes precedence over \texttt{default-days}. \\
\texttt{semver-patch-days} & Delay for patch SemVer updates. Takes precedence over \texttt{default-days}. \\
\texttt{include} & Dependencies the cooldown applies to (wildcards allowed). \\
\texttt{exclude} & Dependencies omitted from the cooldown (wildcards allowed). Takes precedence over \texttt{include}. \\
\bottomrule
\end{tabular}
\end{table}

For each cooldown field, we extract the configured delay values and the \texttt{include} and \texttt{exclude} lists shown in the table. All measurements come from the observation cutoff of 2026-04-30.

To clarify whether adopters rely on \texttt{default-days} alone or also use the SemVer-level delays and the \texttt{include} and \texttt{exclude} lists, we examine how each key is set. For \texttt{default-days} and each of the three SemVer keys \texttt{semver-major-days}, \texttt{semver-minor-days}, and \texttt{semver-patch-days}, we report the percentage of the ecosystems within repositories that set it, and for the numeric values the median, the most common value, and the maximum over those that set it. To remove the influence of repositories that set several cooldowns in the same ecosystem, up to 22 in our data, the unit of analysis is an ecosystem within a repository, and its value is the minimum across the cooldowns of that ecosystem. We examine differences in values across the cooldowns within the same repository in a later sub-analysis.

We also describe \texttt{default-days} by ecosystem. The unit is an ecosystem within a repository, represented by its minimum \texttt{default-days}. To avoid unstable summaries based on very few cases, we report the median for ecosystems in which at least five repositories set the value.

For the exclude entries, we also read the pull requests and commits that introduced them and analyze the stated motivations.

An adopter may also adjust its configuration while keeping the cooldown feature enabled.
We analyze these subsequent changes at the level of an ecosystem within a repository. For each field whose value is in days, namely \texttt{default-days}, \texttt{semver-major-days}, \texttt{semver-minor-days}, and \texttt{semver-patch-days}, we count additions and removals based on whether any cooldown in that ecosystem of the repository sets it. An addition is a commit at which the number of cooldowns setting the field in that ecosystem rises from zero to one or more, and a removal is the reverse. The settings at adoption and reintroduction serve as baselines, so keys already present at those points are not counted as additions. We count additions and removals only between consecutive snapshots in which the ecosystem retains cooldown, so a key that disappears with the cooldown or its parent \texttt{updates} entry is not counted as a removal. We count increases and decreases from the value itself. Because one ecosystem may hold several cooldowns, we use the minimum across the cooldowns that set a field as the representative value. For example, if the same ecosystem has two cooldowns that both set \texttt{default-days}, we use the smaller of the two values. For \texttt{include} and \texttt{exclude}, which have no numeric value, we count only additions and removals. We report the frequency of each kind of change.

In addition to the aggregation by ecosystem within a repository, we examine repositories that hold multiple cooldowns with different values.
We count adopter repositories with at least two cooldowns that differ in \texttt{default-days} or any of the SemVer values.

\section{Results}
\label{sec:results}

\subsection{Dataset Overview}

At the cutoff, the population covers 28 distinct package ecosystems, and a repository declares a median of 2 ecosystems and up to 17. Over the window, 1{,}455 commits edited a Dependabot configuration file and the Dependabot bot opened 95{,}516 version-bump pull requests.

% The remainder of this section reports the adoption and abandonment of cooldown and their motivations for RQ1, adopter characteristics for RQ2, and the customization of cooldown configuration values for RQ3.
\begin{table}[tbp]
\caption{Repository counts and counts of ecosystems within repositories by adoption category.}
\label{tab:dataset-scale}
\centering
\begin{tabular}{@{}lrr@{}}
\toprule
\textbf{Category} & \textbf{Repository} & \textbf{Ecosystem within repo.} \\
\midrule
Adopter                                  &   135 &   252 \\
\quad Continuous adopter                 &   132 &   243 \\
\quad Re-adopted                         &     3 &     8 \\
\quad Abandoned                          &     0 &     1 \\
Non-adopter                              & 1{,}327 & 2{,}298 \\
\midrule
\textbf{Total}                           & \textbf{1{,}462} & \textbf{2{,}550} \\
\bottomrule
\end{tabular}
\end{table}

\subsection{RQ1: Adoption and Abandonment}
\label{sec:rq1-results}

For the cutoff distribution, the final analysis population consists of 1,462 repositories, and 2,550 ecosystems within repositories have an \texttt{updates} entry present at the cutoff. Table~\ref{tab:dataset-scale} gives the four-category breakdown of Section~\ref{sec:study-design} for repositories and for these ecosystems. The repositories split into 135 adopters and 1,327 non-adopters. The ecosystems split into 252 adopters and 2,298 non-adopters, and the 252 adopters comprise 243 continuous adopters, 8 re-adopted adopters, and 1 abandoned adopter. A re-adopted adopter dropped its cooldown to zero and restored it at a later commit, and an abandoned adopter removed cooldown and did not restore it by the cutoff. The 243 continuous and 8 re-adopted ecosystems, 251 in total, hold a cooldown at the cutoff, while the 1 abandoned ecosystem does not. At the repository level no adopter counts as abandoned, and every adopter keeps a cooldown in at least one ecosystem at the cutoff. No adopter withdrew cooldown entirely within the observation window, and when an adopter removed it the removal was confined to some of its ecosystems, which suggests that adopters do not reject the feature itself but keep it while narrowing the scope in which they use it.

\begin{figure}[tbp]
\centering
\includegraphics[width=\columnwidth]{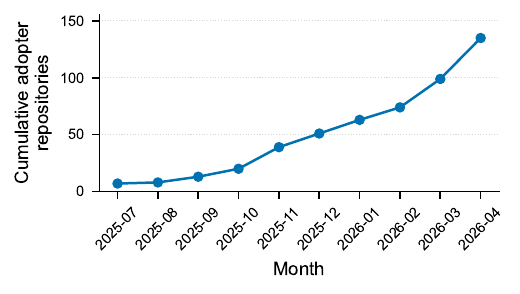}
\caption{Cumulative repository-level adoption over the observation window}
\label{fig:rq1-adoption-timeline}
\end{figure}

For the repository-level adoption timeline, Figure~\ref{fig:rq1-adoption-timeline} shows the monthly cumulative count of adopter repositories. The count rises monotonically throughout the observation window, and the largest monthly increases were in 2026-03 and 2026-04 near the cutoff. This increase is consistent with a feature still early in its diffusion rather than one that has reached saturation, but the 2026-03-31 axios incident may also have prompted some adoptions. Because this study does not causally separate responses to individual incidents, we treat the acceleration near the cutoff as an observed pattern that may include such incident-driven adoption.

Table~\ref{tab:eco-adoption} reports adoption rates by ecosystem. Among the five ecosystems declared by at least 100 repositories, pip has the highest rate at 13.6\%. The mixed-effects model for all 28 ecosystems, which accounts for observations clustered within repositories, completed without a convergence warning or a singular fit. Its omnibus likelihood-ratio test found no significant ecosystem effect ($\chi^2=17.99$, $df=27$, $p=0.904$). However, quasi-complete separation caused by ecosystems with no observed adopters made more than half of the individual coefficients diverge, so the omnibus result should be interpreted with this estimation limitation in mind. In the GEE sensitivity analysis of the 11 ecosystems declared by at least 30 repositories, none of the estimable individual terms was significant (minimum $p=0.057$). The term for composer could not be estimated because it had no observed adopters. This does not establish that ecosystems never differ, but we found no robust evidence of ecosystem-level differences in cooldown adoption.

\begin{table}[tbp]
\caption{Cooldown adoption rate per package ecosystem at the observation cutoff, over the repositories declaring each ecosystem. The 17 ecosystems declared by fewer than 30 repositories form the Others row.}
\label{tab:eco-adoption}
\centering
\begin{tabular}{lrrr}
\toprule
\textbf{Ecosystem} & \textbf{Adopter} & \textbf{Total} & \textbf{Rate} \\
\midrule
github-actions & 116 & 1{,}200 &  9.7\% \\
npm            &  38 &   361 & 10.5\% \\
gomod          &  15 &   210 &  7.1\% \\
pip            &  22 &   162 & 13.6\% \\
docker         &  10 &   116 &  8.6\% \\
cargo          &   6 &    81 &  7.4\% \\
maven          &   7 &    71 &  9.9\% \\
bundler        &  12 &    67 & 17.9\% \\
gradle         &   7 &    64 & 10.9\% \\
composer       &   0 &    47 &  0.0\% \\
nuget          &   2 &    47 &  4.3\% \\
Others         &  16 &   124 & 12.9\% \\
\midrule
\textbf{Total} & \textbf{251} & \textbf{2{,}550} & \textbf{9.8\%} \\
\bottomrule
\end{tabular}
\end{table}

We label the 135 repository-level first-time adoption events with the motivation categories defined in Section~\ref{sec:study-design}. Cohen's kappa between the two raters was 0.90, which is almost perfect under the guidelines of Landis and Koch. Table~\ref{tab:rq1-motivation} reports the distribution with a representative example per category.

Security (Linter) is tied with Unknown as the largest category, which shows that many cooldown additions were made in response to warnings from security linters such as zizmor. By contrast, repositories under the Maintenance label adopted cooldown to reduce pull request noise rather than for security. For the 43 Unknown events, the repositories likely had some intent behind the introduction, but it did not appear on GitHub. Among the 92 events with an identified motivation, 75 are security-only: 43 Security (Linter) and 32 Security (No-Linter). Another 8 events combine security with maintenance concerns. Thus, security is present in 83 identified events, but most of this signal comes from security-only adoption rather than from ambiguous mixed cases. Within the 75 security-only events, 43 (57.3\%) are triggered by a security linter warning such as zizmor. This suggests that security linter warnings can act as an adoption trigger or compliance pressure, making cooldown adoption partly tool-mediated rather than purely a maintainer-initiated security decision.
\begin{table*}[tbp]
\caption{Distribution of adoption motivations over the 135 first-time adoption events, with a representative quote per category. PR denotes pull request.}
\label{tab:rq1-motivation}
\centering
\setlength{\tabcolsep}{4pt}
\begin{tabularx}{\textwidth}{@{}lrr>{\raggedright\arraybackslash}X@{}}
\toprule
\textbf{Motivation} & \textbf{n} & \textbf{\%} & \textbf{Representative quote} \\
\midrule
Security (Linter)        & 43 & 31.9\% & ci: fix issues flagged by Zizmor \\
Security (No-Linter)     & 32 & 23.7\% & Prevent some supply-chain attack vectors \\
Security and maintenance &  8 &  5.9\% & improve dependabot configuration: more stability \\
Maintenance              &  6 &  4.4\% & deps(dependabot): less noisy PRs \\
Template tracking        &  3 &  2.2\% & Synchronize shared configuration \\
Unknown                  & 43 & 31.9\% & Add cooldown to dependabot \\
\midrule
Total                    & 135 & - & \\
\bottomrule
\end{tabularx}
\end{table*}

We identified three repository-level removal events, all in Homebrew repositories. Across these events, cooldown was removed from six ecosystem configurations. The maintainers were investigating why Dependabot pull requests were not opening, and later configuration updates restored cooldown in all three repositories. We also found three ecosystem-level removals in repositories that retained cooldown elsewhere. The nodejs/node github-actions entry is the only one that remained without cooldown at the cutoff. Its maintainer removed the setting after documentation and a continuous integration failure indicated that github-actions did not support it. The other two cases involved docker in future-architect/vuls and Homebrew/brew. In both, cooldown was removed when the unsupported setting prevented Dependabot from running and was restored after docker support became available. The observed removals therefore reflect operational or compatibility problems rather than lasting rejection of cooldown.

\begin{rqbox}{RQ1 Summary}
Answering RQ1, 135 of the 1{,}462 repositories adopted cooldown, with adoption continuing to grow and known motivations centered on security, often through linter warnings. No adopter abandoned cooldown at the repository level, and the few observed removals were generally temporary responses to operational or compatibility problems.
\end{rqbox}
\FloatBarrier

\subsection{RQ2: Adopter Characteristics}
\label{sec:rq2-results}

We first use univariable comparisons to describe how adopters and non-adopters differ in each observed attribute. We then estimate a multivariable model to assess whether each attribute is associated with adoption after adjusting for the other attributes.

After BH adjustment, Tables~\ref{tab:rq2-continuous-stats} and~\ref{tab:rq2-binary-stats} show that adopters have significantly more stars, contributors, dependencies, and bot pull requests, as well as significantly higher rates of organization ownership and \path|SECURITY.md| presence. Among repositories with at least one merged Dependabot pull request, the 124 adopters also have significantly shorter merge times than the 1{,}034 non-adopters. The bot pull request count has a medium effect, while the other significant differences are small or negligible. Repository age and auto-merge usage show no statistically significant differences after BH adjustment. These univariable relationships may change when correlated attributes are considered together.

\begin{table}[tbp]
\caption{Adopter and non-adopter comparison for the six continuous variables at the observation cutoff. The sample includes 135 adopters and 1{,}327 non-adopters, except for dependencies (135 and 1{,}326) and merge time (124 and 1{,}034). Cells are group medians, with repository age in years and merge time in hours. Bold indicates a p-value that is significant at $q=0.05$ after Benjamini--Hochberg (BH) adjustment.}
\label{tab:rq2-continuous-stats}
\centering
\setlength{\tabcolsep}{2pt}
\begin{tabular}{@{}lrrrrl@{}}
\toprule
\textbf{Variable} & \textbf{Adopter} & \textbf{Non-ad.} & \textbf{BH-adj. $p$} & \textbf{$\delta$} & \textbf{Mag.} \\
\midrule
Stars               & 11{,}464 & 9{,}051 & \textbf{0.024} & $+0.12$ & negligible \\
Repository age      & 11.06 & 10.87 & 0.054 & $+0.10$ & negligible \\
Contributors        & 360 & 179 & \textbf{$<$0.001} & $+0.32$ & small \\
Bot pull request count & 54 & 19 & \textbf{$<$0.001} & $+0.36$ & medium \\
Merge time          & 10.87 & 19.17 & \textbf{0.009} & $-0.15$ & small \\
Dependencies        & 187 & 91.5 & \textbf{0.002} & $+0.17$ & small \\
\bottomrule
\end{tabular}
\end{table}

\begin{table}[tbp]
\caption{Adopter and non-adopter comparison for the three binary variables at the observation cutoff. Bold indicates a p-value that is significant at $q=0.05$ after Benjamini--Hochberg (BH) adjustment.}
\label{tab:rq2-binary-stats}
\centering
\setlength{\tabcolsep}{3pt}
\begin{tabular}{@{}lrrrrl@{}}
\toprule
\textbf{Variable} & \textbf{Adopter} & \textbf{Non-ad.} & \textbf{BH-adj. $p$} & \textbf{$V$} & \textbf{Mag.} \\
\midrule
Organization owner & 88.1\% & 68.7\% & \textbf{$<$0.001} & 0.12 & small \\
Auto-merge          & 20.0\% & 13.4\% & 0.054 & 0.05 & negligible \\
\texttt{SECURITY.md} & 54.8\% & 34.4\% & \textbf{$<$0.001} & 0.12 & small \\
\bottomrule
\end{tabular}
\end{table}

Table~\ref{tab:rq2-meta-lang} shows relatively high adoption rates for Python, JavaScript, and Ruby, and lower rates for Go, TypeScript, and Java. Fisher's exact test on the full ungrouped language distribution is significant ($p=0.009$), with a small Cram\'er's V of 0.268.

The pre-GA configuration edit rate is also higher among adopters. Across 115 adopters and 1{,}174 non-adopters, the medians are 1.45 and 0.84 edits per year, respectively ($p=4.7 \times 10^{-7}$, Cliff's $\delta=+0.28$, 95\% CI [$+0.18$, $+0.39$]). This small difference predates cooldown adoption.

\begin{table}[tbp]
\caption{Primary language of each repository for adopters and non-adopters. Languages with fewer than 10 adopter repositories are grouped as Other.}
\label{tab:rq2-meta-lang}
\centering
\setlength{\tabcolsep}{4pt}
\begin{tabular}{lrrrr}
\toprule
\textbf{Category} & \textbf{Adopter} & \textbf{Non-adopter} & \textbf{Total} & \textbf{Rate} \\
\midrule
Python & 37 & 204 & 241 & 15.4\% \\
Go & 17 & 208 & 225 & 7.6\% \\
TypeScript & 13 & 140 & 153 & 8.5\% \\
JavaScript & 18 & 117 & 135 & 13.3\% \\
Java & 10 & 101 & 111 & 9.0\% \\
Ruby & 11 & 50 & 61 & 18.0\% \\
Other ($<10$ adopter repositories) & 29 & 507 & 536 & 5.4\% \\
\bottomrule
\end{tabular}
\end{table}

\begin{table}[tbp]
\caption{Multivariable logistic regression of cooldown adoption on repository attributes and the primary language, reporting odds ratios (ORs) and 95\% confidence intervals (CIs). Continuous variables are log-transformed and standardized, so an OR is per one standard deviation increase. Language ORs are relative to Python. The p-values are unadjusted. The model covers 1{,}461 repositories with 135 adopters and 1{,}326 non-adopters, with a McFadden pseudo $R^2$ of 0.13 and an apparent area under the receiver operating characteristic curve (AUC) of 0.77.}
\label{tab:rq2-multivariate-model}
\centering
\setlength{\tabcolsep}{4pt}
\begin{tabular}{@{}lrrr@{}}
\toprule
\textbf{Variable} & \textbf{OR} & \textbf{95\% CI} & \textbf{$p$} \\
\midrule
Contributors        & 1.45 & [1.12, 1.87] & 0.005 \\
Bot pull request count & 1.85 & [1.44, 2.37] & $<$0.001 \\
Organization owner  & 1.79 & [1.00, 3.23] & 0.051 \\
\texttt{SECURITY.md} presence & 1.68 & [1.14, 2.47] & 0.008 \\
Stars               & 0.93 & [0.76, 1.14] & 0.511 \\
Repository age      & 1.03 & [0.82, 1.29] & 0.817 \\
Dependencies        & 0.90 & [0.69, 1.17] & 0.431 \\
Auto-merge          & 0.91 & [0.55, 1.49] & 0.702 \\
\midrule
\multicolumn{4}{@{}l}{\textit{Primary language, reference Python}} \\
Go          & 0.36 & [0.19, 0.69] & 0.002 \\
Java        & 0.33 & [0.15, 0.74] & 0.007 \\
JavaScript  & 0.78 & [0.39, 1.54] & 0.471 \\
Ruby        & 1.16 & [0.51, 2.61] & 0.725 \\
TypeScript  & 0.45 & [0.21, 0.96] & 0.038 \\
Other       & 0.30 & [0.18, 0.52] & $<$0.001 \\
\bottomrule
\end{tabular}
\end{table}

The largest absolute Spearman correlation observed among the repository attributes was 0.47. The maximum VIF in the main model was 2.1, below the screening criterion of 5. These diagnostics provided no indication that collinearity substantially destabilized the coefficient estimates. The exploratory multivariable model estimates in Table~\ref{tab:rq2-multivariate-model} suggest positive associations of adoption with the number of contributors, bot pull request count, and the presence of a \path|SECURITY.md| file, with adjusted odds ratios of 1.45, 1.85, and 1.68. The estimate for organization ownership is borderline (OR 1.79, $p=0.051$), while stars, repository age, dependencies, and auto-merge show no clear evidence of adjusted associations. Relative to Python, the estimates also suggest lower adoption odds for Go (OR 0.36), Java (OR 0.33), and TypeScript (OR 0.45), but no clear differences for JavaScript or Ruby. These exploratory observational associations do not establish causality.

Two sensitivity models add merge time and the pre-GA edit rate separately because each is defined only for a subset. The merge-time model covers 1{,}158 repositories and finds no clear association for the added variable (OR 0.81, 95\% CI [0.65, 1.02]). The pre-GA edit-rate model covers 1{,}289 repositories, including 115 adopters, and likewise finds no adjusted association (OR 1.15, 95\% CI [0.94, 1.40]). In both models, the main associations and language results remain stable. Thus, merge time differs significantly in the univariable comparison after BH adjustment but has no clear association after accounting for the other explanatory variables.

\begin{rqbox}{RQ2 Summary}
Answering RQ2, the exploratory adjusted estimates suggest associations of adoption with the number of contributors, bot pull request count, and \path|SECURITY.md| presence, as well as lower adoption odds for Go, Java, and TypeScript repositories than for Python repositories.
\end{rqbox}
\FloatBarrier

\subsection{RQ3: Configuration Choices}
\label{sec:rq3-results}

In RQ3, we analyze cooldown configuration values observed at the cutoff among the 135 early adopter repositories, rather than treating the main snapshot as the initial configuration at the adoption event.
The configuration distribution is based on 251 ecosystems within those repositories that still hold cooldown at the cutoff.
We examine post-adoption changes separately only where the git history lets us observe transitions from the initial setting to later settings.
We organize the results into three parts: the set rate and value distribution of each key, value changes after adoption, and the comparison across the cooldowns within the same repository.

Table~\ref{tab:rq3-param-usage} reports the set rate, median, most common value, and maximum of each cooldown key observed at the cutoff over the 251 ecosystems within repositories that hold cooldown. \texttt{default-days} is set in 97.2\% of them, while each of the three SemVer-level keys stays below 10\%. The \texttt{include} and \texttt{exclude} lists appear in only 1.2\% and 2.0\%. Most thus rely on \texttt{default-days} alone at the cutoff. They maintain minimal customization, keeping a single \texttt{default-days} value rather than tuning the delay at each SemVer level or for individual packages. The remaining 7 that do not set \texttt{default-days} each carry a delay in a SemVer-level key, so all 251 of them set a delay in at least one key. None of them is an ecosystem that carries no delay and resolves to zero days.
\begin{table}[tbp]
\caption{Set rate, median, most common value, and maximum of each cooldown key over the 251 ecosystems within repositories with a cooldown at the observation cutoff.}
\label{tab:rq3-param-usage}
\centering
\setlength{\tabcolsep}{4pt}
\begin{tabular}{lrrrr}
\toprule
\textbf{Key} & \textbf{Set rate} & \textbf{Median} & \textbf{Mode} & \textbf{Max} \\
\midrule
\texttt{default-days}      & 97.2\% (244) & 7 & 7 & 42 \\
\texttt{semver-major-days} &  9.2\% (23)  & 21 & 30 & 90 \\
\texttt{semver-minor-days} &  8.0\% (20)  & 7 & 7 & 30 \\
\texttt{semver-patch-days} &  7.6\% (19)  & 3 & 3 & 90 \\
\texttt{include}           &  1.2\% (3)   & - & - & - \\
\texttt{exclude}           &  2.0\% (5)   & - & - & - \\
\bottomrule
\end{tabular}
\end{table}

We first look at the distribution of the \texttt{default-days} value. Among the 244 ecosystems within repositories that set it, the most common value is 7 days, which occurs in 157 of them, or 64.3\%. The values range from 2 to 42 days but concentrate strongly on 7 days. Broken down by ecosystem, all 10 ecosystems in which at least five repositories set the value have a median of 7 days.

The \texttt{include} list is set in 3 ecosystems across 2 repositories, all specifying the wildcard *. All 3 were present when cooldown was adopted.
The \texttt{exclude} list is set in 5 ecosystems across 5 repositories. One was present at adoption, and the other 4 were added later.
In these few cases, the pull requests and commits that introduced each exclude entry suggest three motivations.
The first is the exclusion of packages maintained by the same organization, which carry lower supply chain risk than third-party dependencies.
The second is the exclusion of security-critical libraries, such as cryptographic packages, for which rapid patch adoption is needed.
The third is the exclusion of an action that was blocking continuous integration at the time the exclude entry was introduced.

Changes to the settings after adoption are few. Table~\ref{tab:rq3-value-changes} is the history-based part of RQ3 and reports the counts by direction. \texttt{default-days} is raised 6 times and newly added 4 times, with no decrease or removal, and the 6 increases all move it from 3 or 4 days to 7. The three SemVer-level keys change only by removal, 6, 5, and 5 times. Each of these keys was set in the commit that adopted cooldown and was removed only at a later commit. The observed post-adoption changes fold the separate delays for each update type, all set at adoption, into a single \texttt{default-days}, or raise that \texttt{default-days} toward 7 days. The \texttt{include} and \texttt{exclude} lists are each added 4 times, with no removal while cooldown remains present. The 4 \texttt{include} additions all set the wildcard * in 4 ecosystems across 3 repositories owned by the Homebrew organization. The corresponding \texttt{updates} entries were later deleted and reintroduced without \texttt{include}, so these are not the 3 \texttt{include} settings observed at the cutoff.
\begin{table}[tbp]
\caption{Direction of configuration changes per ecosystem within a repository while the cooldown feature stays enabled.}
\label{tab:rq3-value-changes}
\centering
\setlength{\tabcolsep}{2.5pt}
\begin{tabular}{@{}lrrrr@{}}
\toprule
\textbf{Key} & \textbf{Increased} & \textbf{Decreased} & \textbf{Added} & \textbf{Removed} \\
\midrule
\texttt{default-days}      & 6 & 0 & 4 & 0 \\
\texttt{semver-major-days} & 0 & 0 & 0 & 6 \\
\texttt{semver-minor-days} & 0 & 0 & 0 & 5 \\
\texttt{semver-patch-days} & 0 & 0 & 0 & 5 \\
\texttt{include}           & - & - & 4 & 0 \\
\texttt{exclude}           & - & - & 4 & 0 \\
\bottomrule
\end{tabular}
\end{table}

At the cutoff, 90 of the adopter repositories hold two or more cooldowns. Comparing the values across the cooldowns of each repository, the values set in both settings differ in only 3 repositories. \texttt{default-days} differs in 2 of them, and the \texttt{semver-major-days}, \texttt{semver-minor-days}, and \texttt{semver-patch-days} keys differ in 1.

The 3 repositories with differing values illustrate context-specific policies. Two vary \texttt{default-days} across ecosystems. One gives Go dependencies used for vulnerability data a shorter delay than docker and github-actions, while the other gives its large npm dependencies a longer delay than Go and github-actions. The remaining repository differentiates two settings within npm, using shorter SemVer delays for production and test dependencies and longer delays for the remaining development dependencies. These cases show that the rare departures from a common repository-wide value reflect update priorities rather than arbitrary variation.

\begin{rqbox}{RQ3 Summary}
Answering RQ3, most observed cooldown configurations use only a simple \texttt{default-days} setting, commonly 7 days. Post-adoption changes were rare and generally simplified separate delays into a single delay or moved \texttt{default-days} to 7 days.
\end{rqbox}

\section{Discussion}
\label{sec:discussion}

\subsection{Interpretation of Results}

Early adopters use cooldown more as a simple repository-level delay than as a highly customized scheduling mechanism, and security linters mediate many adoption decisions. This interpretation requires caution because 43 Unknown events account for 31.9\% of the 135 first-time adoptions. The claim that security motivations dominate therefore applies most directly to the 92 events with identified motivations. Even if every Unknown event were non-security motivated, however, security would still be present in 83 events, or 61.5\% of all adoptions, so it remains a major adoption driver.

The rising adoption curve and absence of repository-level abandonment are consistent with an early diffusion phase in which some adopters continue to find value in the feature. The acceleration near the cutoff may partly reflect reactions to the axios incident, whose contribution this study does not isolate. The observed removals suggest that maintainers may temporarily disable cooldown when it is unsupported or prevents Dependabot from running, rather than because they reject the feature itself.

\subsection{Implications for Tool Designers}

The preference for a simple \texttt{default-days} value suggests that low configuration burden matters. The common 7-day value offers an empirical reference point among early adopters, not evidence of an optimal delay. Linters and configuration tools can support adoption by checking ecosystem support, suggesting simple candidate settings, and reserving fine-grained controls for dependencies with clear priorities. Unsupported settings should also fail clearly without stopping Dependabot as a whole, so maintainers can distinguish an invalid cooldown from a general service failure.

After our observation window, GitHub made a three-day cooldown the default for Dependabot version updates without requiring explicit configuration, while leaving security updates immediate and allowing repositories to customize or opt out of the default~\citep{githubDependabotDefaultCooldown2026}. This change is consistent with our implication that a useful baseline should impose little configuration burden. However, the three-day platform default differs from the seven-day value most common among early adopters in our data, and our study does not establish either period as optimal. Future studies should distinguish acceptance of the default, explicit customization, and opt-out behavior when evaluating cooldown use.

\subsection{Implications for Open-Source Software Maintainers}

RQ1 identified three adoption events introduced through shared-configuration tracking (Table~\ref{tab:rq1-motivation}). It also identified removal cases in which unsupported cooldown settings prevented Dependabot from running (Section~\ref{sec:rq1-results}). These findings suggest that projects inheriting cooldown through a shared configuration should verify ecosystem support and ensure that the inherited policy fits their update workflow.

The RQ2 multivariable model shows a positive association between the number of Dependabot pull requests and cooldown adoption (OR 1.85, Table~\ref{tab:rq2-multivariate-model}). Repositories that receive many dependency update pull requests are therefore plausible candidates for evaluating cooldown.

In RQ3, \texttt{exclude} appeared in only five ecosystems across five repositories (Table~\ref{tab:rq3-param-usage}). The cases examined in Section~\ref{sec:rq3-results} involved packages maintained by the same organization, dependencies requiring rapid security updates, and dependencies needed to keep continuous integration working. These observations suggest that maintainers should identify dependencies that must not wait behind the general delay, while ordinary third-party dependencies can retain a detection window~\citep{chinthanet2021lags, cogo2021downgrades}.

\subsection{Threats to Validity}

\subsubsection{Construct Validity}

Our abandonment measure (RQ1) counts only cases in which the cooldown setting is removed while the Dependabot configuration file remains in place. This is a clear signal of cooldown rejection while Dependabot itself continues to be used. Cases in which the Dependabot configuration file is deleted entirely are excluded, because such deletions indicate discontinuation of Dependabot rather than rejection of cooldown specifically, even though cooldown dissatisfaction may be one of several reasons behind them. The abandonment rate should therefore be interpreted as a lower bound on cooldown loss.

The presence of a cooldown setting in the Dependabot configuration file does not necessarily mean that the feature is actively used in practice. A configuration may have been added by a template or a project generator without an explicit decision by the maintainer to enable the cooldown. To partially mitigate this threat, we examine the git history of configuration files and inspect adoption motivations for all 135 first-time adoption events using available commit messages and related Issues and Pull Requests. The \texttt{exclude} list within cooldown removes specified dependencies from the cooldown so that they are updated immediately. A repository counted as an ``adopter'' may therefore apply little real delay to some of its dependencies. We report the usage of \texttt{exclude} lists separately to characterize this nuance.
Our auto-merge measurement for RQ2 captures only pull requests on which auto-merge was enabled at the pull-request level through GitHub's native auto-merge feature. It does not detect Dependabot pull requests that are merged automatically by third-party mechanisms such as GitHub Actions workflows or GitHub Apps (e.g., Mergify). \citet{he2023automating} report that 8.17\% of Dependabot pull requests are merged by third-party auto-merge implementations. Pull requests merged solely through these third-party paths are recorded as having auto-merge disabled in our data, so our auto-merge enablement rate should be interpreted as a lower bound on the prevalence of automated Dependabot pull request merging.

For multiplicity, we treat the nine RQ2 univariable comparisons as one family because they are parallel tests of adopter and non-adopter differences, and we control the false discovery rate within this family using the Benjamini--Hochberg procedure. We do not adjust the multivariable regression coefficients for multiplicity because the model is intended to estimate mutually adjusted associations exploratorily rather than to conduct simultaneous confirmatory tests of individual coefficients. Its $p$-values and 95\% confidence intervals therefore provide no simultaneous error-rate control, so we interpret coefficient-level patterns cautiously and do not base conclusions on borderline results alone.

The RQ2 multivariable logistic regression reports associations from cross-sectional observational data and does not establish causality.
The Dependabot pull request count and merge time are measured over the full observation window, so adopter measurements may be affected by cooldown adoption itself. These metrics therefore do not purely represent pre-adoption repository characteristics, and the observed group differences should not be interpreted causally.
The sensitivity analyses have a further limitation. The analyses that add merge time and the pre-GA edit rate are restricted to repositories with a merged bot pull request and to repositories that had the configuration file at GA, so they narrow the population rather than drawing a random subset. When we compare these sensitivity results with the main model, we therefore account for the difference in the covered set of repositories.

Our dependency count for RQ2 has several limitations. The dependency graph does not separate direct from transitive dependencies, and we count them together. The graph also expands to different depths across ecosystems. For npm and Go it reaches transitive dependencies, whereas for many ecosystems it returns a list close to the direct dependencies, so the count depends on the ecosystem. For some ecosystems the graph returns no packages at all, reporting none for Docker and capturing few dependencies for Gradle and Maven, so the counts for these ecosystems are near zero. The count therefore approximates the dependency update burden rather than giving an exact dependency count. Because the count is ecosystem-dependent in this way, comparing the repository total between adopters and non-adopters could be confounded by an imbalance in ecosystem composition. However, the comparison of adoption rates across ecosystems shows no evidence of a difference even after accounting for within-repository correlation, so any large distortion from ecosystem composition appears unlikely.

\subsubsection{Internal Validity}

Our repository list is obtained from Gitstar Ranking at a single point in time, and repositories whose star counts or rankings change between the scrape of Gitstar and the calls to the GitHub REST API may be captured inconsistently. We record the collection timestamp to make the snapshot reproducible.

For RQ1's qualitative analysis, not all adopter repositories have Issues or Pull Requests that document the rationale for enabling the cooldown, which may bias the sample toward projects with more transparent development processes.

\subsubsection{External Validity}

Our study is limited to public repositories hosted on GitHub. Dependabot is a GitHub-native service and does not operate on other platforms such as GitLab or Bitbucket, so our findings describe the Dependabot user population but cannot be generalized to dependency management practices on other platforms.

We also restrict scope to Dependabot. Renovate and cooldown options built into package managers (npm, pnpm, Yarn, Bun, uv) are not analyzed here, and adoption patterns observed for Dependabot may not transfer to those alternatives.

Because Dependabot's cooldown feature was introduced in July 2025, our observation window spans approximately ten months (2025-07-01 to 2026-04-30), which limits the temporal scope of RQ3 and means that abandonments occurring after the window are not captured in our RQ1 analysis of observed cooldown removals and abandonment.

Finally, we draw our population from the 10,000 repositories with the most stars listed in Gitstar Ranking, so our results may not generalize to smaller or less popular projects, private repositories, or enterprise repositories, whose security practices and tooling choices may differ from those in the community of popular open-source projects~\citep{kalliamvakou2014promises}. Our findings should therefore be interpreted as early adoption patterns for Dependabot cooldown in popular and visible open-source GitHub repositories, rather than as representative evidence for open-source projects in general.

\section{Conclusion}
\label{sec:conclusion}

This paper provides a systematic characterization of early Dependabot cooldown adoption. Through an exploratory study of popular public open-source repositories on GitHub, we examine adoption and abandonment, adopter characteristics, and configuration choices. Together, these analyses establish an empirical baseline for understanding how cooldown is used before its effectiveness is evaluated.

Adoption remains limited but is growing, with 135 adopters among 1{,}462 repositories and no repository-level abandonment during the observation window. Security concerns dominate among events with identified motivations, while linter warnings also trigger adoption. After adjustment, adoption remains associated with more contributors, more bot pull requests, and a \path|SECURITY.md| file. Most configurations apply one general delay, commonly 7 days. Early adopters therefore use cooldown primarily as a low-configuration safeguard rather than a highly customized scheduling policy.

These findings suggest that tool designers should favor simple defaults, validate ecosystem support, and report invalid configurations clearly. Repositories receiving many dependency update pull requests are plausible candidates for evaluating cooldown, but the common 7-day setting should not be interpreted as optimal. Our study provides foundational evidence about early adoption and use, not evidence that cooldown prevents supply chain incidents.

Our evidence is limited to popular public GitHub repositories and an observation window of approximately ten months. Future work should compare configured delays with attack and detection timelines, track explicit customization and opt-out behavior after Dependabot's three-day default, and extend the analysis to Renovate and package-manager cooldowns. These studies should ultimately determine which delay periods balance security benefits against the cost of postponing legitimate updates.

\section*{Declarations}

\subsection*{Funding}
This work was supported by JST BOOST, Japan Grant Number JPMJBS2423, JSPS KAKENHI Nos. JP24K14895 and JP26K21197.

\subsection*{Competing interests}
The authors have no relevant financial or non-financial interests to disclose.

\subsection*{Ethics approval}
Not applicable. This study analyzes publicly available data from open-source software repositories and does not involve direct interaction with human participants.

\subsection*{Informed consent}
Not applicable.

\subsection*{Data availability}
The replication package is available at \url{https://doi.org/10.5281/zenodo.21331117}.

\bibliographystyle{spbasic}      % basic style, author-year citations {ieeetr}
\typeout{}
\bibliography{reference}   % name your BibTeX data base

\end{document}